\documentclass[letterpaper]{article} 
\usepackage{aaai25}  
\usepackage{times}  
\usepackage{helvet}  
\usepackage{courier}  
\usepackage[hyphens]{url}  
\usepackage{graphicx} 
\usepackage{multirow}
\usepackage{tcolorbox}
\usepackage{amsmath}
\usepackage{algorithm}
\usepackage{algorithmic}
\usepackage{array}
\usepackage{mmstyles}
\usepackage{xcolor}
\usepackage{subcaption}
\usepackage{mwe}

\usepackage{natbib}  
\usepackage{caption} 
\usepackage{algorithm}
\usepackage{algorithmic}

\usepackage{color}

\usepackage{array}
\makeatletter
\def\hlinew#1{%
  \noalign{\ifnum0=`}\fi\hrule \@height #1 \futurelet
   \reserved@a\@xhline}
\makeatother

\usepackage{newfloat}
\usepackage{listings}
\DeclareCaptionStyle{ruled}{labelfont=normalfont,labelsep=colon,strut=off} 
\floatstyle{ruled}
\newfloat{listing}{tb}{lst}{}
\floatname{listing}{Listing}
\title{Semantic Convergence: Harmonizing Recommender Systems via Two-Stage Alignment and Behavioral Semantic Tokenization}
\author{
    Guanghan Li\textsuperscript{\rm 1}, Xun Zhang\textsuperscript{\rm 1}, Yufei Zhang\textsuperscript{\rm 1}, Yifan Yin\textsuperscript{\rm 1}, Guojun Yin\textsuperscript{\rm 1}\thanks{Corresponding author.}, Wei Lin\textsuperscript{\rm 1}
}
\affiliations{
    \textsuperscript{\rm 1}Meituan, Beijing, China\\
    \{liguanghan02, zhangxun12, zhangyufei08, yinyifan, yinguojun02, linwei31\}@meituan.com
}

\usepackage{bibentry}

\begin{document}

\maketitle

\begin{abstract}
Large language models (LLMs), endowed with exceptional reasoning capabilities, are adept at discerning profound user interests from historical behaviors, thereby presenting a promising avenue for the advancement of recommendation systems. However, a notable discrepancy persists between the sparse collaborative semantics typically found in recommendation systems and the dense token representations within LLMs. 
In our study, we propose a novel framework that harmoniously merges traditional recommendation models with the prowess of LLMs. We initiate this integration by transforming ItemIDs into sequences that align semantically with the LLMs' space, through the proposed \textit{Alignment Tokenization} module. Additionally, we design a series of specialized supervised learning tasks aimed at aligning collaborative signals with the subtleties of natural language semantics. To ensure practical applicability, we optimize online inference by pre-caching the top-K results for each user, reducing latency and improving efficiency.
%
Extensive experimental evidence indicates that our model markedly improves recall metrics and displays remarkable scalability of recommendation systems. 
\end{abstract}

\section{Introduction}

\begin{figure}[!ht]
    \centering  
    \includegraphics[width=0.45\textwidth]{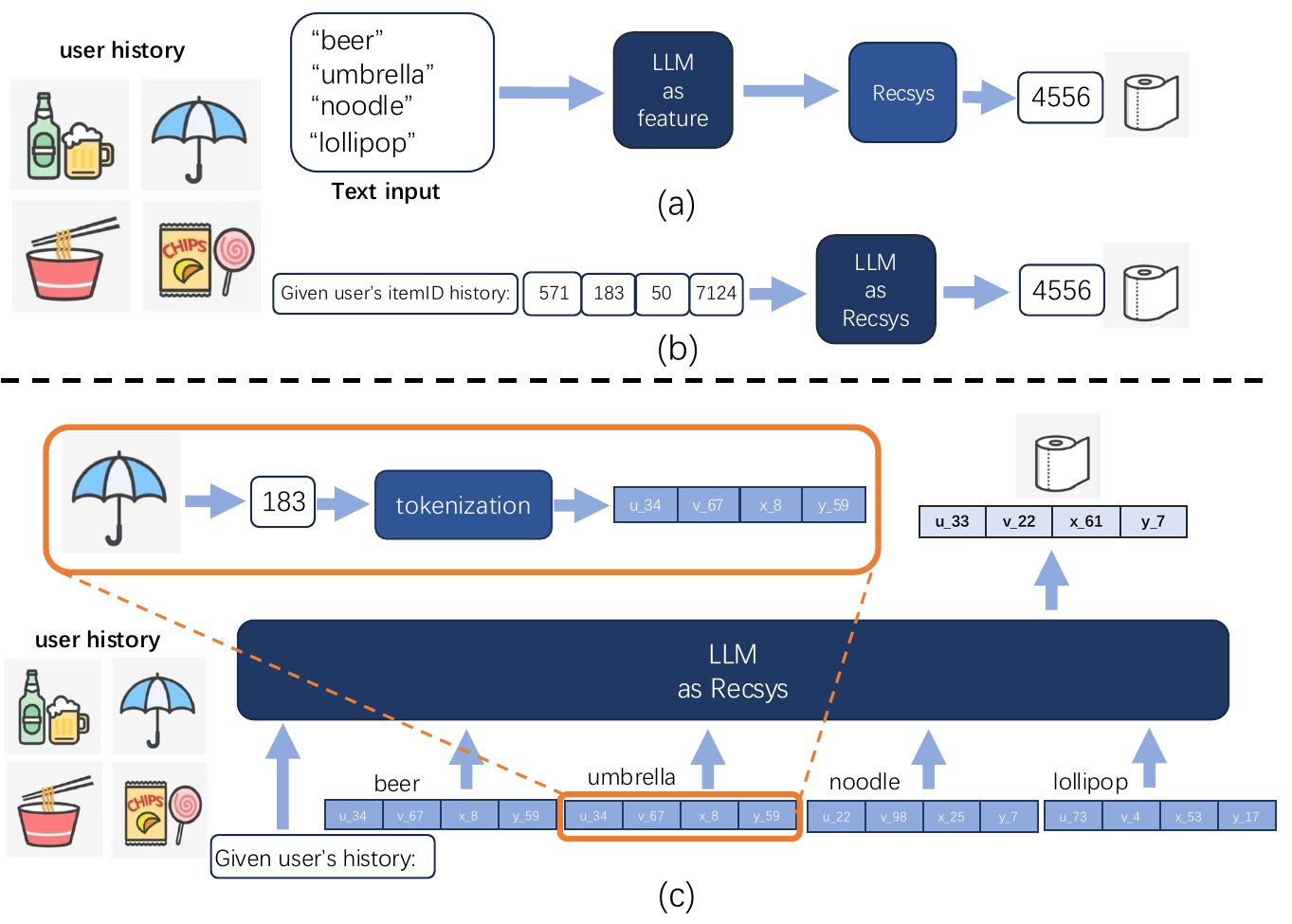} 
    \caption{Illustration of LLM-Based Recommendation Methods. (a) and (b) illustrates existing methods, where either textual information is input into the LLM to provide features for existing recommendation models ~\cite{ren2024representation} in (a), or large-scale item IDs are directly fed into the LLM through instruction fine-tuning, enabling the LLM to directly output recommended item IDs~\cite{li2023e4srec} in (b). (c) represents our method, which can compress large-scale item IDs into a small-scale token representation that is easier for the LLM to process. Furthermore, our method allows the LLM to simultaneously handle both item IDs and textual information, directly outputting the recommended item IDs.}
    \label{fig:overall_figure}
\end{figure}

Recent advancements in Large Language Models (LLMs) technologies, particularly those exemplified by trailblazing models such as GPT-4, have signified a substantial progression in the sphere.
The intersection of these advancements with recommendation technologies has engendered significant intrigue due to the renowned proficiency of LLMs in the area of advanced natural language processing.
This confluence suggests a promising trajectory for augmenting semantic comprehension and behavioral inference within recommendation systems.

Nevertheless, the fusion of LLMs with recommendation systems engenders a unique set of complexities.
The modality of language token representation in LLMs is fundamentally divergent from the myriad of sparse identifiers employed by traditional recommendation algorithms.
This semantic representation disparity precipitates considerable alignment and scalability challenges~\cite{zheng2023adapting}, elements that are crucial for the efficacious implementation of these hybrid systems.
Existing methodologies, inclusive of fine-tuning LLMs on user behavior sequences~\cite{zhang2023collm, li2023e4srec} and the integration of supplementary item identifiers into the LLMs~\cite{zhu2024collaborative, ren2024representation}, has
several limitations as depicted in Figure.~\ref{fig:overall_figure}. Firstly, in existing methods, as shown in (a), the LLM is either used merely as a feature extractor (LLM as feature), where textual information is fed into the LLM to extract semantic features to enhance existing recommendation models. However, this approach may be limited by the incompatibility between the semantic information generated by the LLM and the recommendation models based on behavior information. It also overlooks the LLM's ability to understand user behavior sequences. Alternatively, as illustrated in (b), large-scale item ID sequences are trained into the LLM through instruction fine-tuning. While this method attempts to leverage the LLM's understanding of user behavior sequences, it lacks important textual semantic signals. Moreover, the large-scale item IDs in real industrial environments increase the difficulty of fine-tuning the LLM.

In response to these challenges, as shown in (c), we propose an innovative two-stage alignment framework, conceived to synchronize recommendation models with LLMs via a process of semantic convergence.
Initially, the proposed \textit{Alignment Tokenization} Module is entrusted with the translation of item embeddings from recommendation systems into sequences that are semantically congruent with the representations utilized by LLMs.
This pivotal step serves to bridge the divide between sparse and dense representations, thereby augmenting the comprehension of user interests.
Subsequent to this, we have engineered a series of \textit{Alignment Tasks}, meticulously crafted to further hone the semantic calibration of LLMs.
These tasks strategically leverage the inherent features of recommendation systems, thus enabling LLMs to more astutely discern user interests across a variety of domains.
%
Our framework, which integrates harmoniously with extant systems, has achieved substantial progress in improving recommendation metrics and scalability.

Our contributions can be encapsulated as follows:

\noindent \textbf{1)} A two-stage recommendation system is designed with LLMs, effectively aligning the semantics of both behavior and language signals.

\noindent \textbf{2)} A novel methodology has been devised to map the product ID representation, derived from the recommendation system, to a sequential representation that can undergo further training with any LLM structure or traditional recommendation model.

\noindent \textbf{3)} We propose several fine-tuning tasks for LLM, which includes Sequential alignment, text alignment, and negative sampling strategies. This approach effectively models user interests and enhances the robustness of the task.

\section{Related Works}
\noindent \textbf{Sequential Recommendation.}
The field of Sequential Recommendation is widely applied, with ongoing research dedicated to forecasting consumer preferences based on their historical interactions.
%
%
LSTM~\cite{45881} and GRU4Rec~\cite{hidasi2015session} have demonstrated their proficiency in capturing both long-term dependencies and immediate associations.
Current endeavors are increasingly adopting sophisticated graph convolutional neural network models within recommendation systems like NGCF~\cite{wang2019neural} and lightGCN~\cite{he2020lightgcn}.
Additionally, Transformer-based models, \eg SASRec~\cite{kang2018self}, BERT4Rec~\cite{sun2019bert4rec}, S3-Rec~\cite{zhou2020s3} have garnered attention for their ability to leverage self-attention mechanisms.

\vspace{0.1cm}
\noindent \textbf{LLM in Recsys.}
Incorporating Large Language Models (LLMs) into Recommendation Systems (RS) has become a hot topic in recent research, thanks to the vast knowledge base and superior reasoning abilities of LLMs.
%
%
By enhancing the representation of IDs~\cite{hou2022towards, hua2023index, hou2023learning}, or tweaking the design of training tasks and the foundational model structure~\cite{li2023e4srec,zhang2023collm, lin2023multifacet, zhu2024collaborative, wang2024rethinking, wang2024llm}, LLMs are given recommendation capabilities.
%
%
Some studies also use LLM as a feature enhancer~\cite{lyu2023llm, xi2023towards, di2023retrieval, wei2024llmrec}, but the main recommendation task is still managed by traditional models, which may not fully utilize the reasoning abilities of LLMs.
%
%
Solutions such as NoteLLM~\cite{zhang2024notellm, zhang2024notellm2} and others~\cite{ren2024representation}, which are based on representation learning, or system integration methods~\cite{luo2024integrating}, have been proposed, but they have complex processes and are not easily scalable.
%
%
Our method uses LLM as a recommendation system, fully aligning semantic information with collaborative semantics, and is also compatible with traditional recommendation models, making the framework easily adaptable to any recommendation scheme.

\begin{figure*}[t]
\centering
\includegraphics[width=1.0\textwidth]{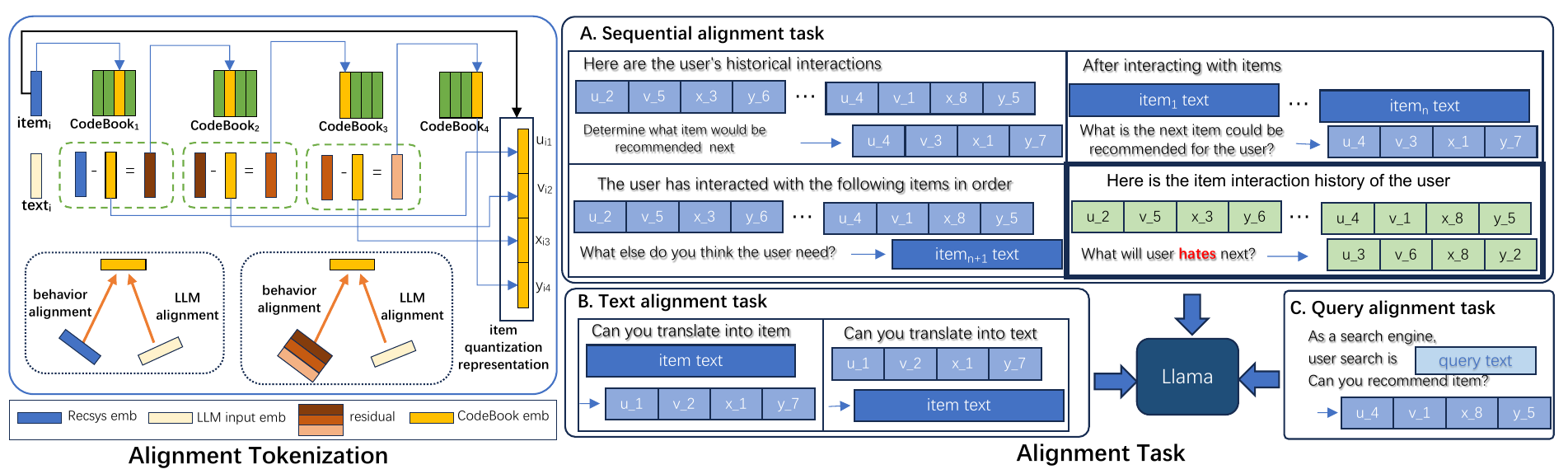} 
\caption{The pipeline of ours two-stage alignment for recommendation. During the Alignment Tokenization phase, our objective is to obtain four token indices—$u$, $v$, $x$, and $y$ for each item. Subsequently, in the Alignment Task process, we utilize these item indices to introduce negative sampling tasks (highlighted by black boxes in the Alignment Task) alongside various positive prediction tasks. These tasks collectively fine-tune the LLM, integrating both behavioral and semantic signals into the model.}
\label{fig:teaser}
\end{figure*}

\section{Method}
\label{sec:method}
In this section, we present our two-stage alignment framework for Large Language Models (LLMs) in recommendation systems, as depicted in Fig.~\ref{fig:teaser}. The first stage, \textbf{Alignment Tokenization}, mitigates the inefficiency of LLM training due to the vast scale of items. This is achieved by mapping items onto a discrete vector set (tokenization). During this process, we introduce an alignment module to better synchronize the tokenization with the LLM's input semantic space. The second stage, \textbf{Alignment Task}, enhances the LLM's ability to predict user interests by incorporating training data that is beneficial to recommendation task into LLM's training process. Concurrently, We pre-cache predictions of LLM to facilitate feasible online usage. These modules will be elaborated further in the subsequent sections.

\subsection{Alignment Tokenization}
\label{subsec:alignment_tokenizer}

To enable LLM to comprehend items, it is necessary to represent items as tokens in the LLM's vocabulary. However, in recommendation scenarios, the number of items is typically vast. Utilizing the original item IDs as LLM tokens would result in increased training sparsity and cost. 
To address this issue, we propose a mapping method that transforms large-scale item space into smaller discrete space. Our approach involves constructing a small-scale discrete index library, that is, the $\textit{CodeBook}_{1}$ to $\textit{CodeBook}_{4}$ on the left side of Fig. \ref{fig:teaser}, where each item is represented by four indices from the CodeBooks. These indices can be shared among items, with more related items sharing greater number of indices. 

To create the CodeBooks, we draw inspiration from the concept of RQ-VAE~\cite{rajput2024recommender}. Firstly, we define a cascaded CodeBooks with $N$ levels, ranging from coarse to fine-grained, with each layer containing $C$ codes. Each item selects an optimal code from each layer, and is thereby represented by $N$ codes.

In training stage, for the first layer of the CodeBooks, We initially cluster the codes into $C$ centers based on a batch of input item embeddings, which serve as the initial code embeddings of the first layer. It is worth noting that LLM have strong semantic understanding capabilities but lack an understanding of user behavior. Based on this, we utilize embeddings derived from DCCF~\cite{ren2023disentangled} trained exclusively on behavior as the item embeddings. Subsequently, the item embeddings will serve as the input for the first layer, identifying the nearest code in the first layer of the CodeBooks to form \textit{embedding pairs} for calculating the distance loss. 
For the subsequent CodeBooks layers, we use the residual of \textit{embedding pairs} in the previous layer as input of current layer. The $C$ initial cluster centers and \textit{embedding pairs} of the current layer are both generated based on this input. Clearly, the distance loss for the current layer is also calculated from the \textit{embedding pairs}. Consequently, $N$ distance losses are summed. This process demonstrates that the CodeBook approximates the item's embedding by iteratively approximating the residuals generated from previous layers. This is shown as \textit{behavior alignment} on the left side of Fig.~\ref{fig:teaser}

This process which can be understood as the tokenization of LLM leverages user behavior to map item IDs into a smaller space. This space, however, is independent of LLM, potentially resulting in misalignments between the semantic space of tokenization and input of LLM. To address this challenge, we have implemented an \textit{LLM alignment} loss, which is depicted in the Fig.~\ref{fig:teaser} on the left. This \textit{LLM alignment} mechanism operates by imposing penalties on the discrepancies between LLM embedding and the codes, thereby ensuring that each code layer is effectively synchronized with the semantic space by the LLM. The LLM embedding are represented by the average pooling of the LLM input layer embedding through the titles and descriptions of items.

The whole training process of this cascaded CodeBooks can be described as follows:
\begin{equation}
    E_{i}^{n} = \left| E^{(n-1)}_i - B^{(n-1)}_{c^{*}} \right| \quad \text{if} \ n \geq 1, E_{i}^{0} = \text{Emb}(\text{r}_i)
\end{equation}

\begin{equation}
    c^{*} = \arg\min_c \text{Dist}(E^n_i, B^n_c)
\end{equation}

\begin{equation}
    \label{eq:alignment}
    L_{i}^{B} = \sum_{n=0}^{N} \text{Dist}(E^n_i, B^n_{c^{*}}), L_{i}^{L} = \sum_{n=0}^{N} \text{Dist}(B^n_{c^{*}}, E^{\text{LLM}}_i)
\end{equation}

\begin{equation}
    L_{i} = L_{i}^{B} + L_{i}^{L}
\end{equation}


where $Emb(r_i)$ denotes the $i_{th}$ item embedding from DCCF~\cite{ren2023disentangled}, i.e. \textit{Recsys emb} in Fig.~\ref{fig:teaser}. $E_{i}^{n}$ is the input to the $n_{th}$ layer of CodeBooks, and $B^n_c$ refers to the representation of $c_{th}$ code in $n_{th}$ layer of CodeBooks. $Dist()$ denotes a vector distance function, such as cosine distance, and $E^{LLM}_i$ represents the embedding of the $i_{th}$ item at the LLM input layer. In our experiments, we chose llama-7B~\cite{llama1} as the LLM. The loss function consists of \textit{Behavior Alignment} loss $L_{i}^{B}$ and \textit{LLM Alignment} loss $L_{i}^{L}$ as depicted on the left side of Fig.~\ref{fig:teaser}

After the CodeBooks has been trained, each item searches for the nearest code in each layer of CodeBooks according to the training process, ultimately representing each item with $N$ codes. In our experiments, $N=4$ and $C=256$. These codes are subsequently incorporated into the LLM's vocabulary to represent the items during the fine-tuning process.

Note that, unlike previous work~\cite{rajput2024recommender}, our method overlook the auto-encoding process. This distinction arises from our objective of constructing discrete index library, whereas the auto-regressive encoder is primarily designed for generation tasks. Pursuing an unrelated task objective would compromise the final accuracy. We conducted a quantitative analysis in Tab.~\ref{tab:ATM}.

\subsection{Alignment Task}
\label{subsec:alignment_task}

After obtaining the item quantization representation for each item through our Alignment Tokenization, the subsequent task involves fine-tuning the LLM using user interaction and text descriptions. To prevent training instability arising from substantial dimensional differences of embeddings between CodeBooks and LLM, We utilize only the code indices of each item rather than their embeddings. Consequently, the newly introduced tokens representing items remain in an untrained state until the LLM undergoes fine-tuning.
Following prior work~\cite{zheng2023adapting}, as depicted on the right side of Fig.~\ref{fig:teaser}, several fine-tuning tasks are defined including \textit{Sequential alignment task}, \textit{Text alignment task}, \textit{Query alignment task}. The following are examples for each task:
\tcbset{
  colback=gray!5!white,
  colframe=gray!75!black,
  fonttitle=\bfseries
}

\begin{tcolorbox}
\textbf{A. Sequential alignment task}

[prompt]: Here is the item interaction history of the user: \textless$item_i$\textgreater,..., what to recommend to the user next?

[label]: \textless$item_j$\textgreater

\vspace{1em}

\textbf{B. Text alignment task}

[prompt]: An item is called $title$ and described as $description$, can you tell me which item it is?

[label]: \textless$item_j$\textgreater

\vspace{1em}

\textbf{C. Query alignment task}

[prompt]: You meet a user's query: $query$. Please respond to this user by selecting an appropriate item

[label]: \textless$item_j$\textgreater
\end{tcolorbox}
where the \textless$item_i$\textgreater \ and \textless$item_j$\textgreater \ refer to token index of the item derived from Alignment Tokenization, the $title$ and $description$ represent the item's title and textual description, respectively, and the $query$ denotes user's review. For all tasks, [prompt] are utilized as the training inputs for LLM, while [label] serves as the training targets. These tasks are integrated into training set to fine-tuning LLM with cross-entropy loss, a widely used loss for training LLMs. Llama-7B~\cite{llama1} is selected as our LLM.


The user interaction item sequence is a vital component of the training data. However, the number of interacted items per user is typically limited. Utilizing only interacted items as training data can lead to over-fitting and introduce sample selection bias~\cite{ma2018entire} due to insufficient data.
To address this challenge, as depicted in the Fig.~\ref{fig:teaser}, the green area is enclosed by black box, we introduced negative sampling strategy for user behavior to align with the behavior generalization capabilities of traditional recommendation models. For scenarios without negative samples, we randomly sample items that the user has not interacted with. Based on the negative samples of user behaviors, we have added a ``negative behavior task" in the Sequential alignment task. This task is used to represent items that users are unlikely to interact with in the current context. The negative samples are incorporated into a new set of negative prompt expressions, as a example illustrated below:
\begin{tcolorbox}
\textbf{Negative sampling task}

[prompt]: Here is the item interaction history of the user: \textless$item_i$\textgreater,..., what will user \textbf{hate} next?

[label]: \textless$item_j$\textgreater
\end{tcolorbox}
The data corresponding to these negative prompt expressions are used along with other tasks as training data for LLM fine-tuning. The efficacy of our negative sampling strategy is illustrated in Tab.~\ref{tab:ATM}.

It is noteworthy that in industrial recommendation systems, various training acceleration techniques, such as Low-Rank Adaptation (LoRA)~\cite{hu2021lora} and FlashAttention~\cite{dao2022flashattention}, can be utilized to expedite the training process. Furthermore, incremental training can help in reducing training data. Nevertheless, it is important to acknowledge that despite the implementation of these acceleration methods, the training speed of LLM remains slower than that of traditional recommendation models. This issue is a prevalent challenge faced across the this field.

\begin{figure}[t]
    \centering  
    \includegraphics[width=0.5\textwidth]{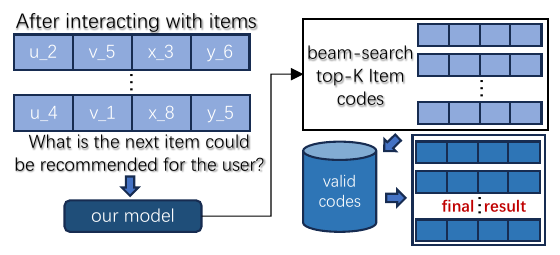} 
    \caption{Inference Stage. We cache valid codes in predictions for each user during the online inference phase.}
    \label{fig:inference}
\end{figure}

\subsection{Inference}
\label{subsec:inference}
Following supervised fine-tuning, The prompt ``\textit{The user has interacted with \textless$item_i$\textgreater,... in chronological order. Can you predict the next
possible item that the user may expect?}'' which mentioned in LC-Rec~\cite{zheng2023adapting} is used for inference. However, due to the large number of parameters in the LLM, the inference process incurs significant computational costs and exhibits slow inference speed, rendering it unsuitable for real industrial system. To address this limitation, as depicted on the Fig.~\ref{fig:inference}, we pre-cache the top-$K$ item codes for each user, generated through beam search inference within the LLM. The \textit{valid codes} refers to the collection of item codes corresponding to all possible items obtained from the CodeBooks. We only cache valid item codes. During the online inference phase, retrieval of the relevant user's cached data suffices. 

It is important to note that when new items are introduced into the item pool, it is unnecessary to retrain LLM and CodeBooks. The token representation of items can be derived through inference during our alignment tokenization stage. Despite the fact that new items have not been trained in the LLM, these items have acquired meaningful representations during CodeBook's inference, thereby mitigating the cold start problem for new items.

\begin{table*}[ht]
	\centering
	\begin{tabular}{c|c|m{1.2cm}<{\centering}|m{1.5cm}<{\centering}|m{1.1cm}<{\centering}|m{1.1cm}<{\centering}|m{1.1cm}<{\centering}|m{1.2cm}<{\centering}|m{1.1cm}<{\centering}|m{1.3cm}<{\centering}}
		\hlinew{1.1pt}
            Dataset & Metrics & SASRec & BERT4Rec & FMLP  & DCCF  & P5 & LC-Rec  & Ours & Improv.\\
            \hline 
		\multirow{5}{*}{Games} & HR@1 & 0.0013 & 0.0136 & 0.0154 & 0.0197 & 0.0158 & 0.0279 & \textbf{0.0323} & +15.77\% \\ 
            & HR@5 & 0.0296 & 0.0468 & 0.0496 & 0.0578 & 0.0495 & 0.0755 & \textbf{0.0858} & +13.64\%\\
            & HR@10 & 0.0576 & 0.0773 & 0.0806 & 0.0882 & 0.0789 & 0.1167 & \textbf{0.1274} & +9.17\%\\
            & NDCG@5 & 0.0153 & 0.0302 & 0.0325 & 0.0389 & 0.0329 & 0.0518 & \textbf{0.0591} & +14.09\%\\
            & NDCG@10 & 0.0243 & 0.0400 & 0.0424 & 0.0486 & 0.0423 & 0.0651 & \textbf{0.0724} & +11.21\%\\
            \hline
		\multirow{5}{*}{Arts} & HR@1 & 0.0015 & 0.0184 & 0.0372 & 0.0339 & 0.0487 & 0.0639 & \textbf{0.0676} & +5.79\%\\ 
            & HR@5 & 0.0423 & 0.0467 & 0.0689 & 0.0696 & 0.0705 & 0.0999 & \textbf{0.1041} & +4.20\%\\
            & HR@10 & 0.0643 & 0.0655 & 0.0922 & 0.0954 & 0.0876 & 0.1237 & \textbf{0.1289} & +4.20\%\\
            & NDCG@5 & 0.0225 & 0.0329 & 0.0534 & 0.0520 & 0.0597 & 0.0822 & \textbf{0.0863} & +4.99\%\\
            & NDCG@10 & 0.0295 & 0.0389 & 0.0608 & 0.0602 & 0.0652 & 0.0899 & \textbf{0.0942} & +4.78\%\\
            \hline
            \multirow{5}{*}{Instruments} & HR@1 & 0.0002 & 0.0396 & 0.0527 & 0.0440 & 0.0595 & 0.0708 & \textbf{0.0731} & +3.25\%\\ 
            & HR@5 & 0.0523 & 0.0625 & 0.0794 & 0.0742 & 0.0834 & 0.0985 & \textbf{0.1022} & +3.76\%\\
            & HR@10 & 0.0714 & 0.0797 & 0.1016 & 0.0929 & 0.1023 & 0.1225 & \textbf{0.1268} & +3.51\%\\
            & NDCG@5 & 0.0276 & 0.0516 & 0.0663 & 0.0595 & 0.0717 & 0.0848 & \textbf{0.0879} & +3.66\%\\
            & NDCG@10 & 0.0338 & 0.0571 & 0.0734 & 0.0656 & 0.0777 & 0.0925 & \textbf{0.0958} & +3.57\%\\
            \hlinew{1.1pt}
	\end{tabular}
\caption{Comparative analysis of metrics across various models and datasets, with the best results emphasized in bold. As shown in the table, the ``LC-Rec" is the second-best method after our approach. Therefore, ``Improv." indicates the relative improvement of our method compared to ``LC-Rec".}
\label{tab:result}
\end{table*}

\section{Experiment}
\label{sec:exp}


\subsection{Experiment Setup}
\label{subsec:exp_setup}
\subsubsection{Datasets}
We evaluated our proposed method by utilizing the ``Games", ``Arts" and ``Instruments" datasets from the Amazon review dataset~\cite{ni2019justifying}. These datasets encompass user interactions with items, with each item accompanied by a title and description. To mitigate the impact of long-tail data on training, we define the maximum sequence length for user interactions as 20. Furthermore, samples with a concatenated prompt length exceeding 2048 were excluded. The statistical details of these publicly available datasets are presented in Table.~\ref{tab:stat_datasets}. 

\begin{table}
	\centering
	\begin{tabular}{c|c|c|c|c}
		\hlinew{1.1pt}
            Dataset & Users & Items & Len avg & Long user\\
            \hline 
		  Games & 50546  & 16859 & 8.962 & 4.80\% \\ 
            \hline  
            Arts & 45141 & 20956 & 8.658 & 4.34\% \\
            \hline
            Instruments & 24772 & 9922 & 8.322 & 3.48\%\\
            \hlinew{1.1pt} 
	\end{tabular}
\caption{Statistics of the datasets. In addition to the number of ``Users" and ``Items", ``Len avg" denotes the average length of user sequences, while ``Long user" denotes the proportion of users whose sequence length exceeds 20, relative to the total number of users.}
\label{tab:stat_datasets}
\end{table}

\subsubsection{Baselines}
For comparison purposes, we chose several representative recommendation models as baselines: 
\begin{itemize}
\item \textbf{SASRec~\cite{kang2018self}}: A self-attention driven sequence recommendation model captures long-term dependencies within user behavior sequences.
\item \textbf{BERT4Rec~\cite{sun2019bert4rec}}: The sequence recommendation method captures patterns in users' behavior by utilizing BERT's bidirectional encoding.
\item \textbf{FMLP~\cite{zhou2022filter}}: The model employs learnable filters to enhance the sequence data and utilizes the Fast Fourier Transform to automatically filter out noise.
\item \textbf{DCCF~\cite{ren2023disentangled}}: Self-supervised contrastive learning was employed to achieve intention decoupling and enhances the representations of users and items.
\item \textbf{P5~\cite{geng2022recommendation}}: Various recommendation tasks were integrated into a unified model. We implement it based on this code~\citep{github_LLM_RecSys_ID}. 
\item \textbf{LC-Rec~\cite{zheng2023adapting}}: A quantization method for item indexing was designed, which enhances recommendation capabilities of the LLM through fine-tuning.
\end{itemize}

In all methods, we extract the first $S$-1 actions from user behavior sequence of length $S$ to serve as input prompt and label in training set. In the evaluation process, the user's last interacted item is designated as the label, with the preceding actions as input prompts.


\subsubsection{Metrics}
We employed two widely recognized metrics to assess the performances: top-$K$ Hit Ratio (HR) and top-$K$ Normalized Discounted Cumulative Gain (NDCG). 
In our paper, we utilize HR@K and NDCG@K to represent these metrics, and the parameter $K$ is set to 1, 5, and 10.

\subsection{Experiment Results}
\label{subsec:results}
We conducted a comparison between our method and several baseline models. 
The results are presented in Tab.~\ref{tab:result}. It is evident that the LLM-based approach outperforms traditional methods across all three datasets. This superiority can be attributed to the LLMs, which encompasses a wealth of world knowledge and enhances recommendation effectiveness by incorporating information such as behavior or text. Moreover, our proposed method exhibits improvements over the LLM-based approach, LC-Rec~\cite{zheng2023adapting} on all three datasets, as shown in the last column of Tab.~\ref{tab:result}. This primarily owe to the two-stage alignment that we specifically designed for the recommendation task. 

Our results indicate that the improvement achieved by our method on the \textit{Games} dataset is more significant compared to the \textit{Arts} and \textit{Instruments} datasets. This discrepancy may be attributed to longer average sequence length and higher proportion of long-sequence users in the \textit{Games} dataset, as illustrated in Table.~\ref{tab:stat_datasets}. The richer the user behavior, the more significant performance enhancement of LLM. It suggests that LLMs with robust semantic parsing capabilities possess a superior ability to learn from multi-modal behavior.

Additionally, we found that the gains from comparing fewer recommendation results are greater than those from comparing more results, \eg, the gain in HR@5 is greater than that in HR@10. This maybe because, the LLM only has one positive label token per sample during training, which results in only the top-1 token being penalized by loss function. For this reason, the top-ranked recommendation results benefiting more from our approach.

\subsection{Ablation Study}
\label{subsec:abl}
To demonstrate the benefits of each feature of our proposed method, considering that LC-Rec~\cite{zheng2023adapting} is the best-performing method among baselines, we compare the results with the variants by removing each newly added cue compared to LC-Rec~\cite{zheng2023adapting}, \ie, 1) ``ED", our tokenization incorporates an Encoder-Decoder similar to RQ-VAE~\cite{rajput2024recommender}, 2) ``w/o LA", our method without the \textit{LLM alignment} loss $L^L$ in eq.~\eqref{eq:alignment}, 3) ``w/o NS", our method without negtive sampling, as shown in Tab.~\ref{tab:ATM}.

Our method shows significant improvement over ``ED", indicating that the encoder-decoder module used for generative tasks is not suitable for our clustering representation task. The penalty function designed for generative capabilities can weaken our representation ability. Furthermore, our method shows a slight improvement compared to ``w/o AM", this is because the alignment loss helps pre-align with LLM input space in our tokenization stage, facilitating more efficient convergence of the LLM during fine-tuning phase. The slight improvement in performance may be attributed to LLM's rich general knowledge, which eases convergence challenges. Lastly, our method shows significant improvement over ``w/o NS", thanks to our negative sampling strategy that adds a large amount of effective data, enhances generalizability, and alleviates sample selection bias.

\begin{table}
	\centering
	\begin{tabular}{c|c|c|c|c}
		\hlinew{1.1pt}
             & HR@5 & HR@10 & NG@5 & NG@10\\
            \hline 
		ED & 0.0817 & 0.1197 & 0.0570 & 0.0693 \\
            \hline 
		w/o LA & 0.0836 & 0.1264 & 0.0586 & 0.0723 \\
            \hline 
            w/o NS & 0.0791 & 0.1195 & 0.0548 & 0.0677 \\
            \hline 
		Ours & 0.0858 & 0.1274 & 0.0591 & 0.0724 \\
            \hlinew{1.1pt}
	\end{tabular}
\caption{Ablation study in the Games dataset. ``NG'' denotes NDCG. It showcases the effect of every modification we made based on LC-Rec~\cite{zheng2023adapting}.}
\label{tab:ATM}
\end{table}

\noindent \textbf{Different negative sampling ratios.}
Table.~\ref{tab:NSR} illustrates the evaluation of different negative sampling ratios and their performance. It is evident that as the negative sampling ratio increases, the performance gradually improves, albeit at a diminishing rate when the negative sampling ratio reaches 1:4. This is attributed to our negative sampling strategy, which has enhanced the LLM's generalization ability regarding behaviors and alleviated sample selection bias. 


\begin{table}
	\centering
	\begin{tabular}{c|c|c|c|c}
		\hlinew{1.1pt}
            NSR & HR@5 & HR@10 & NDCG@5 & NDCG@10 \\
            \hline 
		1:0 & 0.0791 & 0.1195 & 0.0548 & 0.0677 \\
            \hline 
		1:1 & 0.0817 & 0.1239 & 0.0559 & 0.0695 \\
            \hline 
		1:2 & 0.0830 & 0.1256 & 0.0574 & 0.0711 \\
            \hline 
		1:3 & 0.0858 & 0.1274 & 0.0591 & 0.0724 \\
            \hline 
		1:4 & 0.0871 & 0.1277 & 0.0606 & 0.0737 \\
            \hline 
		1:5 & 0.0856 & 0.1270 & 0.0597 & 0.0729 \\
            \hlinew{1.1pt}
	\end{tabular}
\caption{Comparison of results with different negative sampling ratios in Games Dataset, where ``NSR" represents the negative sampling ratio. The left and right sides of the colon represent positive and negative samples, respectively}
\label{tab:NSR}
\end{table}


\noindent \textbf{Varying numbers of CodeBooks.}
Table.~\ref{tab:codebooknum} presents the performance for varying numbers of CodeBooks. We observe that when the number of CodeBooks is 2, the collision rate exceeds 10\%. However, when $C$ is 3 or greater, the collision rate becomes negligible. Token sharing among items facilitates more comprehensive token training. Consequently, an excessive number of new words in LLM can result in sparse token training, while too few new words may lead to fewer token representations for individual items, potentially causing inadequate expression. As shown in Table.~\ref{tab:codebooknum}, the optimal performance is achieved when the number of CodeBooks is 4. This scenario represents a balanced state between the number of CodeBooks and tokens, prompting us to select $C=4$ as the number of CodeBooks for our experiments.

\noindent \textbf{Scaling law of recommendation ability.}
To verify whether larger and more advanced LLM exhibit superior performance in our recommendation task, we conducted experimental comparisons across various parameter scales and model versions. The experimental results are presented in Table.~\ref{tab:scaling_law}. The first two models represent different versions with the same parameter scale, whereas the last model feature larger parameter scales. We observed that when utilizing LLMs with similar parameter sizes, more advanced model versions do not necessarily exhibit superior performance. This phenomenon may be attributed to the fact that the advanced models primarily incorporate additional training corpora, which might be significantly different from the recommendation domain. Furthermore, in recommendation tasks, the semantic information tends to be relatively simple, lower-version models are adequate for handling the simple semantic information. However, larger parameter sizes yield improvements in performance. This could be attributed to the fact that models with larger parameters are more effective at learning behavior patterns. Furthermore, the superior performance of larger parameter LLMs implies a scaling law for LLMs in recommendation capabilities. This not only indicates that we can improve recommendation performance by simply increasing the model parameters, but also suggests that LLM-based recommendation models can benefit from the rapid advancements in the LLM domain.

\begin{table}
	\centering
	\begin{tabular}{c|m{0.8cm}<{\centering}|m{0.8cm}<{\centering}|m{1cm}<{\centering}|m{1cm}<{\centering}|m{1cm}<{\centering}|m{1cm}<{\centering}}
		\hlinew{1.1pt} 
            C & CR & T & HR@5 & HR@10 & NG@5 & NG@10 \\
            \hline
		1  & —— & 16859 & 0.0700 & 0.1111 & 0.0470 & 0.0602\\ 
            \hline 
		2 & 10.8\% & 512 & 0.0789 & 0.1198 & 0.0537 & 0.0668\\ 
            \hline 
		3 & 0.05\% & 768 & 0.0804 & 0.1225 & 0.0555 & 0.0690\\ 
            \hline 
		4 & 0.0\% & 1024 & 0.0858 & 0.1274 & 0.0591 & 0.0724\\ 
            \hline 
		5 & 0.0\% & 1280 & 0.0804 & 0.1215 & 0.0554 & 0.0687\\ 
        \hlinew{1.1pt}
	\end{tabular}
\caption{Performance with varying numbers of CodeBooks in Games Dataset. ``C" denotes the number of CodeBooks, indicating that each item is represented by ``C" tokens. ``CR" stands for the collision rate, representing the proportion of items with duplicate token representations among all items. ``T" denotes the number of new words added to the LLM. ``NG" is the abbreviation for NDCG. When C=1, the original item IDs are directly used as token representations, thereby eliminating the possibility of collisions.}
\label{tab:codebooknum}
\end{table}

\begin{table}
	\centering
	\begin{tabular}{c|c|c|c|c}
	\hlinew{1.1pt} 
            model & HR@5 & HR@10 & NG@5 & NG@10 \\
            \hline
		Llama-7B  & 0.0858 & 0.1274 & 0.0591 & 0.0724 \\ 
            \hline 
		Llama-2-7B & 0.0817 & 0.1242 & 0.0562 & 0.0699 \\
            \hline 
		Llama-13B & 0.0910 & 0.1329 & 0.0636 & 0.0771 \\ 
        \hlinew{1.1pt}
	\end{tabular}
\caption{Performance Evaluation Using Different LLMs in Games Dataset. The ``NG" stands for NDCG.}
\label{tab:scaling_law}
\end{table}

\noindent \textbf{Fine-tuning LLMs with different embedding initialization methods.}
When obtaining token representation for each item, we only utilized the ``index" results produced during alignment tokenization phase, omitting the ``embedding" results from this phase. Consequently, the tokens newly added to the LLM were not pre-initialized before fine-tuning. However, the embeddings produced by the CodeBooks during alignment tokenization phase possess semantic significance. Intuitively, initializing LLM's token embeddings with these CodeBooks embeddings might yield better performance. To test this hypothesis, we projected the 32-dimensional embeddings obtained from the DCCF model to the LLM's dimensionality (4096) using a specific mapping module, and then used this projection as the initialization for the LLM's token embeddings. Given the substantial dimensional differences, we omitted experiments with a 1-layer MLP mapping module and instead employed 2-layer and 3-layer MLPs. As illustrated in Table.~\ref{tab:token_init}, our findings indicate that the performances of these mapping modules has declined. This may be attributed to the dimensional and semantic gap between code embeddings and the input embeddings of LLM, which hinders the convergence of the LLM.

\begin{table}
	\centering
	\begin{tabular}{c|c|c|c|c}
		\hlinew{1.1pt} 
            Method & HR@5 & HR@10 & NG@5 & NG@10 \\ 
            \hline 
		2-layer & 0.0721 & 0.1152 & 0.0475 & 0.0614 \\ 
            \hline 
		  3-layer  & 0.0747 & 0.1168 & 0.0501 & 0.0636 \\ 
            \hline
		Ours  & 0.0858 & 0.1274 & 0.0591 & 0.0724 \\
        \hlinew{1.1pt}
	\end{tabular}
\caption{Comparison of token embedding initialization methods for LLM in Games Dataset. ``NG'' denotes NDCG. ``2-layer" and ``3-layer" refer to different modules that map the CodeBook embeddings with 32 dimensions to 4096 dimensions. Specifically, these modules are ``MLP-Relu-MLP", and ``MLP-Relu-MLP-Relu-MLP", respectively. The gradients of these modules and the CodeBook embeddings will be activated during the fine-tuning phase.}
\label{tab:token_init}
\end{table}

\noindent \textbf{Performance comparison using only behavior sequence training.}
In our experimental setup, the first five baseline methods were trained using only user behavior. However, both LC-Rec and our method utilized not only user behavior but also textual information. To exclude the influence of textual information, we compared the performance of all models using only user behavior sequences. We selected the best three methods from baselines, DCCF and P5, as well as LC-Rec which also utilizes textual information, as our baseline comparisons. We conducted experiments using the Games dataset. The experimental results are shown in Table.~\ref{tab:only_seq}. As can be observed, using only user behavior will result in a significant loss of performance. However, our method with using only user behavior significantly outperforms the first two baselines, thanks to the LLM's powerful learning capabilities. Furthermore, our method shows substantial improvement over LC-Rec, which can be attributed to the superior performance of our two-stage alignment method.
\begin{table}
	\centering
	\begin{tabular}{c|c|c|c|c}
		\hlinew{1.1pt} 
            Method & HR@5 & HR@10 & NG@5 & NG@10 \\
            \hline
		DCCF  & 0.0578 & 0.0882 & 0.0389 & 0.0486 \\ 
            \hline 
            P5  & 0.0495 & 0.0789 & 0.0329 & 0.0423 \\ 
            \hline 
	      LC-Rec  & 0.0708 & 0.1044 & 0.0494 & 0.0602 \\
            \hline 
		Ours & 0.0773 & 0.1132 & 0.0537 & 0.0652 \\
            \hline 
		Ours w/ T & 0.0858 & 0.1274 & 0.0591 & 0.0724 \\
        \hlinew{1.1pt}
	\end{tabular}
\caption{Performance Comparison Using Only User Behavior Sequences in Games Dataset. ``NG" stands for the NDCG metric, and ``Ours w/ T" indicates that our method uses complete information. We selected three well-performing methods as our baselines.}
\label{tab:only_seq}
\end{table}

\section{Limitation and Future Work}
While our work has demonstrated outstanding performance, caching results during inference phase may be constrained by storage space and is inadequate for managing frequently changing item pool. Pre-caching vectors reveal an alternative method. However, the beam search decoding method incurs high costs for vector retrieval. Consequently, one of our future endeavors is to enhance efficiency of online inference. Furthermore, the large-scale data inherent in the recommendation domain renders the training costs of LLMs prohibitively high. Thus, improving training efficiency in the recommendation domain is our another focus. Additional, our work only involves text and behavior, we plan to incorporate additional modalities such as images. Lastly, in ablation study on scaling law of recommendation capability, we were constrained by computational resources and thus only conducted experiments with 13B model and below. We plan to allocate more computational resources to the experiments.

\section{Conclusion}
In this paper, we propose a method via Alignment Tokenization and Alignment Task to enhance the recommendation system based LLM. Specifically, for Alignment Tokenization, to alleviate the issues of increased training costs and sparsity caused by large-scale items, we present a method for mapping large-scale items to a smaller-scale index library. Additionally, we introduce LLM alignment loss to pre-align during the tokenization phase, addressing the misalignment between tokenization and input space of LLM. For Alignment Task, to increase data volume and mitigate sample selection bias, we incorporate a negative sampling strategy in the training data of LLM, aligning the traditional recommendation model's ability of generalize user interests. Our method's effectiveness is validated on three datasets.


\clearpage

\bibliography{aaai25}

\end{document}